\documentclass[aps,prl,11pt]{article}
\usepackage{graphicx}
\usepackage{epsfig}
\usepackage{textcomp}
\usepackage{amssymb}
\usepackage{hyperref}

\begin{document}

\title{CF6 White Paper: Gamma-ray Signatures of Ultra High Energy Cosmic Ray Line-of-sight Interactions}

\author{J. Dumm, L. Fortson}
\date{\today}

\maketitle

\section{Origins of UHECRs and Production of Secondary Gamma-rays}
Identifying the origins of Ultra High Energy Cosmic Rays (UHECRs; E $>10^{18}$ eV) has been a long-standing goal in astrophysics.  Cosmic rays arrive at Earth nearly isotropically after being deflected by magnetic fields.  It is likely that most of this deflection occurs within our own galaxy, which has large magnetic fields with respect to the intergalactic magnetic fields (IGMFs).  Only in recent years has there been a claim of an anisotropy at the highest observed energies, linking UHECRs with AGN~\cite{PAO09112007}.

Magnetic field strengths inside galaxies are measured to be about 1 -- 10 $\mu$G using Faraday rotation and Zeeman splitting.  The same magnetic field strengths can be found near the cores of galaxy clusters, while weaker magnetic field strengths of 10 -- 100 nG have been found near the outskirts of galaxy clusters.  The IGMF strength and structure, however, are not well constrained.  Upper limits of $\sim$1 nG exist with some reasonable assumptions on magnetic field correlation length~\cite{Neronov:2009gh}.  Even more constraining upper limits of $\sim$1 fG may apply if AGN efficiently heat the intergalactic medium~\cite{Broderick:2011av, Chang:2011bf, Pfrommer:2011bg}. 

With such small IGMF strengths, if AGN produce UHECRs, these particles may travel largely undeflected for most of their journey to Earth.  These UHECRs will occasionally interact en route with background photons.  The predominant interactions between protons and Extragalactic Background Light (EBL) or cosmic microwave background photons are proton pair production and pion production, both producing secondary gamma-rays, which might be detectable according to a number of studies \cite{Ferrigno:2004am, EK10, Essey:2010er, Prosekin:2012ne, Murase:2011cy, Razzaque:2011jc, Takami:2013gfa}.  

These secondary gamma-rays will appear as an approximately point-like source for the current generation of Imaging Atmospheric Cherenkov Telescopes (IACTs), under reasonable assumptions.  Taking an IGMF strength of 10~fG, deflections are on the order of $0.1^{\circ}$, approximately scaling as
\begin{equation}
\Delta \theta \sim 0.1^{\circ} \left(\frac{B}{10^{-14}\mathrm{G}}\right) \left(\frac{4 \times 10^7~\mathrm{GeV}}{E}\right) \left(\frac{D}{1~\mathrm{Gpc}}\right)^{1/2} \left(\frac{l_c}{1~\mathrm{Mpc}}\right)^{1/2},
\end{equation}
where $D$  is the distance to the source and $l_c$ is the average correlation length of the magnetic domains~\cite{Essey:2010er}.

\section{Status of the Search for Gamma-ray Signatures of UHECR Sources}

Analysis of TeV blazar spectral energy distributions (SEDs) predominantly shows good agreement with electronic synchrotron self-Compton models.  In some cases, an external photon field is required to model the data accurately~\cite{Acciari:2009xz}.  As of yet, no evidence has been established which would require a hadronic component to the jet.  These primary gamma-rays created at the source outshine any secondaries produced en route before EBL absorption is taken into account.  Only for spectral measurements at very high energy or at very high redshift (i.e. at high optical depth) might the flux of UHECR-induced gamma-rays become important since these travel only a fraction of the total distance.  

\begin{figure}[htb!]
\begin{center}
\includegraphics[scale=1.0]{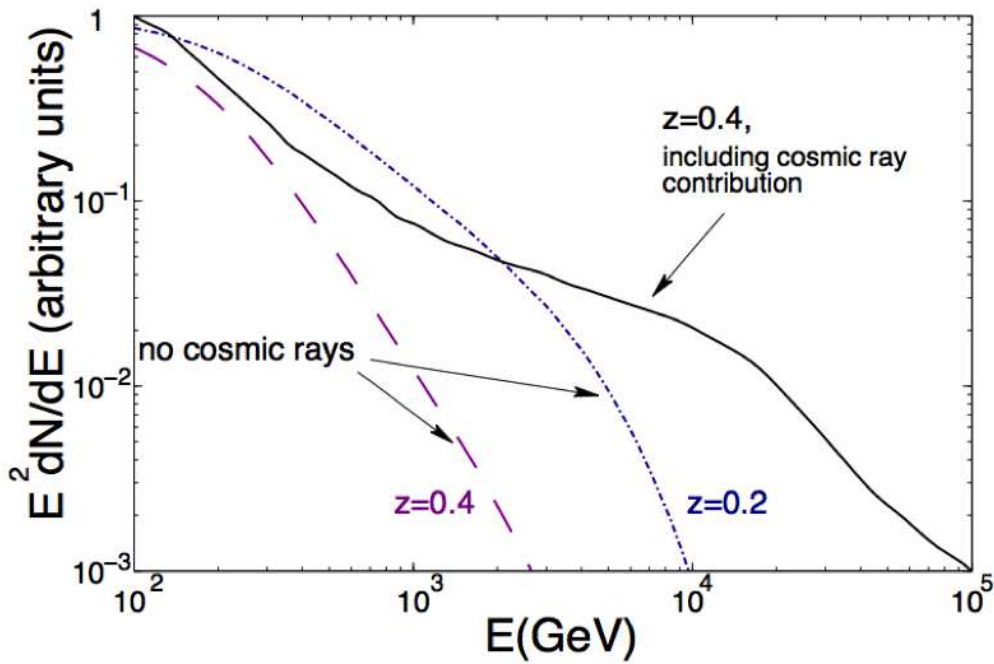}
\caption{\label{fig:spec} Contributions from cosmic ray line-of-sight interactions result in a hardening of the spectrum from distant sources.  Figure taken from~\cite{Prosekin:2012ne}.  }
\end{center}
\end{figure}

Simulations of UHECR interactions have been performed demonstrating the possibility that all observed TeV photons from several distant blazars can be explained by UHECR line-of-sight interactions.  Although the total cosmic ray power and spectrum are  not known, the unknown EBL shape is the dominant factor in determining the spectral shape of the secondary gamma-rays~\cite{Essey:2010er}.  In fact, most indirect EBL measurements using TeV photons ignore the possibility of spectra being modified by UHECR secondaries.  This is an important caveat to any results derived from TeV spectra.   Figure \ref{fig:spec} gives an example of how UHECR secondaries can modify a gamma-ray spectrum if the cosmic ray power is high enough.  

The current generation of IACTs have the sensitivity to such a signature of UHECRs secondaries.  A recent population study looked at 50 spectra of TeV blazars and found an anomaly at the $\sim$4 standard deviation level, showing that the data do not agree with our standard blazar emission and EBL models.  This is evidence that some process, such as photons mixing with axion-like particles, is suppressing pair production~\cite{Horns:2012fx, Meyer:2012sb}.  An alternative explanation is that UHECR-induced gamma-rays are becoming important at high optical depth.  Determining the existence of such an anomaly with higher statistical power and determining its cause should be a focus of current and next-generation TeV telescopes.  

The total deflection of UHECRs may be small enough to motivate gamma-ray observations of the arrival directions of UHECR events.  
VERITAS has performed a dedicated search for gamma-ray emission from two pairs of UHECR events, resulting in upper limits from associated AGN~\cite{Holder:2008uu}. 

\section {The Search for UHECR Origins with CTA}

The Cherenkov Telescope Array can help determine whether or not we see signs of UHECR secondaries in blazar spectra in several ways.  The expected improvement in the point spread function could allow for detection of the slight smearing of UHECR secondary gamma-rays under certain assumptions of the magnetic field.  The improved energy resolution leads to substantially reduced error bars on spectral measurements, greatly enhancing our ability to detect spectral features.  Perhaps most important, small magnetic deflections smear out most time dependence in secondary gamma-ray flux.  This lack of variability may be the key to differentiating between UHECR origins and other possibilities (such as photons mixing with axion-like particles).  In addition to just detecting a larger sample of blazars, the enhanced point-source sensitivity of CTA translates into an ability to determine whether or not a source is flaring with a higher statistical power than current instruments.  Timing signatures of UHECR secondaries are discussed in detail in~\cite{Prosekin:2012ne} while using spectra from CTA observations of known sources to differentiate hadronic from leptonic cascades is discussed in~\cite{Takami:2013gfa}.

The most relevant energy range for detection of UHECR secondary gamma-rays is in the few hundred GeV to tens of TeV range where the EBL absorption becomes strong for many sources.  Recent estimates show that the US contribution of mid-sized telescopes would be particularly important as it improves the point-source sensitivity in this key energy range by a factor of two to three \cite{Jogler:2012ps}.  

\bibliographystyle{unsrt}       
\bibliography{myrefs}           

\end{document}